\documentclass[reprint,aps,pra,superscriptaddress,
]{revtex4-2}

\usepackage[colorlinks=true,linkcolor=blue,citecolor=blue,urlcolor=blue]{hyperref}
\usepackage{amsmath}
\usepackage{braket}
\usepackage{siunitx}
\usepackage[all]{hypcap} 
\usepackage[capitalise]{cleveref}
\usepackage{easyReview}
\usepackage{overpic}
\usepackage{CJKutf8}

\crefname{appendix}{Apd.}{Apds.}
\crefname{table}{Tab.}{Tabs.}
\crefname{section}{Sec.}{Secs.}
\DeclareSIUnit\bar{bar}
\DeclareSIUnit{\mbar}{\milli\bar}

\newcommand*{\sfref}[2]{Fig.~\hyperref[#1]{\ref*{#1}#2}}

\begin{document}

\preprint{APS/123-QED}

\title{Optomechanical Levitation and Control of High Aspect Ratio Silicon Nanorods}

\author{Zhenan Chen~(\begin{CJK*}{UTF8}{gbsn}陈贞安\end{CJK*})}
\thanks{These authors contributed equally to this work.}
\affiliation{%
Department of Physics, King's College London, Strand, London, WC2R 2LS, UK
}%
\email{zhenan.chen@kcl.ac.uk}

\author{Sophie Sanghera}
\thanks{These authors contributed equally to this work.}
\affiliation{%
Department of Physics, King's College London, Strand, London, WC2R 2LS, UK
}%

\author{Yanhui Hu}
\affiliation{%
Department of Physics, King's College London, Strand, London, WC2R 2LS, UK
}%
\author{James Sabin}
\affiliation{%
Department of Physics, King's College London, Strand, London, WC2R 2LS, UK
}%
\author{Maryam Nikkhou}
\affiliation{%
Department of Physics, King's College London, Strand, London, WC2R 2LS, UK
}%
\author{Alex Gee}
\affiliation{%
Kelvin Nanotechnology, University of Glasgow, G12 8LS Glasgow, United Kingdom
}%
\author{David Burt}
\affiliation{%
Kelvin Nanotechnology, University of Glasgow, G12 8LS Glasgow, United Kingdom
}%
\author{Qiongyuan Wu~(\begin{CJK*}{UTF8}{gbsn}吴穹远\end{CJK*})}
\affiliation{%
Department of Physics, King's College London, Strand, London, WC2R 2LS, UK
}%
\author{Joseph Kelly}
\affiliation{%
Department of Physics, King's College London, Strand, London, WC2R 2LS, UK
}%
\author{James Millen}%
\affiliation{%
Department of Physics, King's College London, Strand, London, WC2R 2LS, UK
}%
\affiliation{%
London Centre for Nanotechnology, King's College London, Strand, London, WC2R 2LS, UK
}%
\email{james.millen@kcl.ac.uk}

\date{\today}

\begin{abstract}
Nano- and micro-particles levitated by optical, electrical or magnetic fields are a new frontier in precision sensing and for tests of fundamental physics. For optically levitated anisotropic particles it is possible to control their translation, alignment and rotation. We report on the levitation and characterization of nanofabricated, high uniformity, high aspect ratio, high refractive index silicon cylinders, with diameters as low as 50\,nm and lengths up to 1500\,nm. We are able to tune their oscillation frequencies from 10\,kHz to over 1\,MHz, and exert huge optical torque to generate high rotation rates. These optically levitated silicon nanorods will enable precision torque sensing, and when pushed to smaller sizes, tests of quantum physics through the generation of angular momentum superposition states. 
\end{abstract}

\maketitle

\section{Introduction}\label{sec:introduction}
Light is the most precise and versatile means of controlling matter at the nanoscale. Trapping nano- and micro-objects using focused laser beams \cite{Ashkin1986} has enabled precise control of dielectric particles \cite{Gieseler2012,Gonzalez-Ballestero2021,millen2020}, biological specimens \cite{Wang1997,Bustamante2021}, and other mesoscopic systems \cite{Spesyvtseva2016}, leading to the partial award of a Nobel Prize in Physics in 2018 \cite{Ashkin1986}. In 2010 three proposals suggested that it would be possible to control the motion of levitated nanoparticles at the quantum level \cite{romero2010, chang2010, Barker2010}, a feat which was realised a decade later \cite{Delic2020, Magrini2021, Tebbenjohanns2021}. From this point, various proposals suggest methods for generating macroscopic quantum superposition states of the nanoparticle position \cite{bateman2014, Romero2011} to test quantum mechanics \cite{Millen2020_Quantum, Arndt2014} and even the quantum nature of gravity \cite{Bose2017, Marletto2017}. Nanoparticles are at a scale relevant to technology and biology, and beyond the current state of the art objects for which matter-wave interferometry has been performed \cite{Fein2019, Pedalino2026}. Current experiments report small coherent delocalization of nanoparticles released from optical traps \cite{Rossi2025}, prompting researchers to explore methods to accelerate the expansion of the particle's wavefunction such that some diffracting element can be illuminated for high-mass interferometry \cite{Tomassi2025, Bonvin2024}. 

Another approach to generating macroscopic quantum superpositions is to consider angular, rather than translational, motion. Free angular momentum is quantized, negating the requirement for a diffracting element. An angular momentum superposition would be evidenced by revivals of the initial nanoparticle alignment \cite{stickler2018}, at a characteristic revival time $t = 2\pi I/\hbar$, where $I$ is the moment of inertia of the particle. This quantum revival time can be on the order of a few milliseconds \cite{stickler2018} as opposed to hundreds of milliseconds required for centre-of-mass superpositions \cite{bateman2014, Romero2011}, greatly simplifying decoherence-evading experimental protocols. Such an experiment to observe quantum revivals, and hence infer angular momentum superposition states, requires precise initialization and read-out of the nanoparticle's alignment, which would not be possible with a spherically symmetric object. The protocol would also require a large number of repetitions to accumulate sufficient statistics, necessitating uniform particles. In this work, we optically levitate highly uniform nanofabricated silicon nanorods, which are promising candidates for a future quantum revival experiment.

Access to angular degrees of freedom brings other advantages. When trapped, the angular oscillatory modes (or librational modes) can exhibit frequencies above $1\,$MHz, facilitating deep quantum cooling \cite{Dania2025, Troyer2026}. Simultaneous cavity cooling of all translational and angular modes has been demonstrated \cite{pontin2023}. Angular motion is highly sensitive to applied torques \cite{kuhn2017full, Stickler2021}, with a minimum torque sensitivity (units Nm\,Hz$^{-1/2}$) of $Q_{\rm min}= \sqrt{4k_BTI\Gamma}$, where $T$ is the temperature of the mode and $\Gamma$ is the momentum damping rate of that mode \cite{Hoang2016, Ahn2020}. This has enabled the measurement of torque at the $10^{-27}\,\unit{\newton\metre}$ level. Librational mode frequencies are sensitive to the geometry of the particle, and tiny geometric asymmetries of presumed-spherical particles become noticeable in high-precision experiments \cite{Kamba2023, Arita2023, Ahn_2018}. Angular degrees of freedom therefore provide a versatile platform for precision sensing and for studying classical and quantum rotational dynamics \cite{Stickler2021}.

Optical torque can be applied to levitated particles to make them rotate, through scattering of light carrying angular momentum \cite{kuhn2017full, Stickler2021}. Typically, this coupling requires either birefringence \cite{Arita2023} or shape anisotropy \cite{Ahn_2018}, though spherically symmetric particles can also be made rotating through optical absorption \cite{Monteiro2018}. The polarizability anisotropy of high aspect ratio silicon nanorods  \cite{Donato2019} leads to large optical torque. Nanoparticles in ultra-high vacuum conditions have reached rotation rates of over $1\,\unit{\GHz}$ \cite{Reimann_2018, Jin2021}. Rotating nanoparticles can be used for a range of sensing applications \cite{Ahn2020, Ju2023}, and for probing models of vacuum \cite{Manjavacas2010} and quantum friction \cite{Zhao2012}. Rotation can be used for gyroscopic stabilization \cite{Arita2013,Wang2024} and directly for gyroscopic sensing \cite{Zeng2024,Zieli2024}. 

In this work, we outline the fabrication of high-uniformity silicon nanorods over a range of dimensions and aspect ratios in \cref{sec:fabrication}, and the near-equilibrium theoretical model of their motion in \cref{sec:model}. We demonstrate stable optical levitation of different nanorods from below $1\,\unit{\mbar}$ to near-atmospheric pressures, with tunable motional frequency from $10\,\unit{\kHz}$ to over $1\,\unit{\MHz}$ by adjusting the laser power in \cref{sec:syst}. We characterize the nanorods' motion at different pressures, compare the measurements with analytic model, and extract empirical susceptibilities. In addition, we demonstrate ultra-high optical torques that enable rotation frequencies well above $1\,\unit{\MHz}$ at $100\,\unit{\mbar}$. In \cref{sec:inst}, we discuss the challenges associated with levitating these particles under high-vacuum conditions.

\section{Nanorod Fabrication}\label{sec:fabrication}

\begin{figure}[t]
\includegraphics[width=\linewidth]{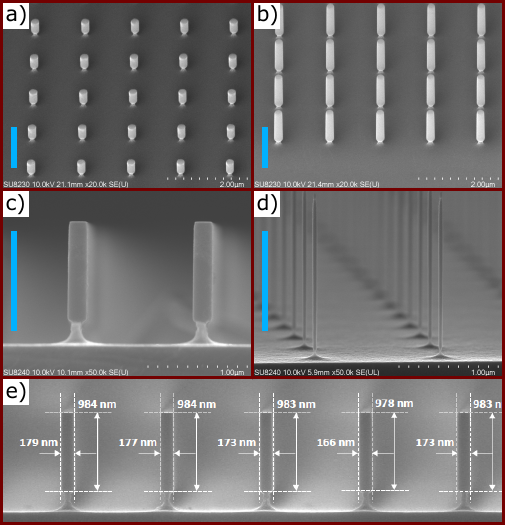}
\caption{FESEM images of silicon nanorods used in this study. Nominal length $\times$ diameter: a) $400\times200\,\unit{\nm}$ (top view), b) $1000\times200\,\unit{\nm}$ (top view), c) $1000\times200\,\unit{\nm}$ (side view), d) $1500\times50\,\unit{\nm}$ (side view) with a small amount of silica etch mask visible at the top. e) Dimension analysis on nominal $1000\times180\,\unit{\nm}$ nanorods (not otherwise used in this study). Based on 30 measurements the dimensions are determined to be $(1024\pm16)\times(174\pm2)\,\unit{\nm}$. Blue bars in a-d) represent a $1\,\unit{\um}$ scale bar. All nanorod samples feature a narrowing at the base to facilitate loading into the optical trap via Laser Induced Acoustic Desorption (LIAD) \cite{nikkhou2021}.}
\label{fig:SEM}
\end{figure}

In this study, we work with silicon nanorods with aspect ratios ranging from 2 to 30. At the trapping wavelength of 1550\,nm, silicon has a high relative permittivity of $\epsilon_R = 12.1$, as compared to $\epsilon_R = 2.1$ for silica (SiO$_2$), the material typically used in studies of levitated optomechanics. This means for a spherical Rayleigh particle under identical optical trapping conditions, the trapping frequencies of silicon are $1.7$ times higher than for silica. Cylindrical particles have a susceptibility anisotropy, as discussed in \cref{sec:syst}, further boosting the trapping frequencies. Both silica \cite{Hebestreit_2018,Zhang_2023} and silicon \cite{Junnemann_2025} have been shown to suffer from absorption related instability, as discussed further in \cref{sec:inst}.  

Field-emission scanning electron microscope (FESEM) images of some of the silicon nanorod samples used in this study are shown in \cref{fig:SEM}.  Silicon nanorods are fabricated from silicon (100) orientation single crystal wafers. Electron beam lithography was used to define an array of SiO$_2$ islands, of target diameters, on a square grid pitch of $1.25\,\unit{\um}$ from Hydrogen silsesquioxane (HSQ) based negative tone resist. The island diameter was transferred to the silicon wafer to form pillar structures using an ICP etch tool (SPTS Omega LPX 200 Rapier) with an anisotropic etch process. A narrowing was introduced to the base of the pillar by switching the plasma etch chemistry to provide an isotropic etch profile. The remaining SiO$_2$ etch mask was removed from the tip of the nanorod using a hydrofluoric acid solution, excepting those with a diameter of 50\,nm due to their fragility.  Nanorod dimensions were characterised using FESEM.

In \sfref{fig:SEM}{e}, we show a uniformity analysis on silicon nanorods of approximate dimension diameter $D = 180\,$nm, length $L = 1000\,$nm. Images are taken using FESEM with a resolution of approximately 5\,nm due to edge blur. The overall height ($L$) is measured, and the diameter at three points (top, middle, bottom), only the middle measurement is shown in the figure.  An analysis of 30 nanorods of this dimension was produced, showing uniformity at the percent level, and a slight tapering from the top to the bottom of the nanorod. For comparison, commercial spherical silica nanoparticles typically have a diameter dispersion of around 15\%.  

\section{Model of optically levitated nanorods}\label{sec:model}

\begin{figure*}[t]
\begin{overpic}[width=\linewidth]{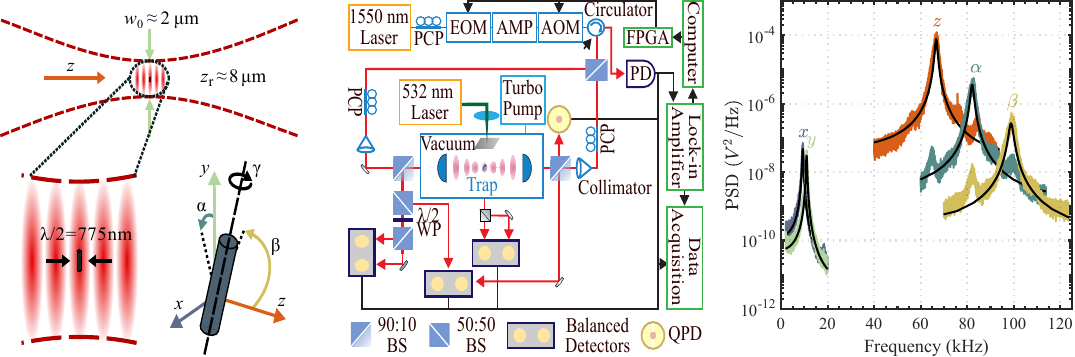}
    \put(0,32){a)}
    \put(30,32){b)}
    \put(67,32){c)}
\end{overpic}
\caption{The experimental setup. a) Illustration of a silicon nanorod held in a counter-propagating standing wave trap, formed by two linearly polarized beams of wavelength 1550\,nm. The translational degrees of freedom are ($x$, $y$, $z$) and the rotational degrees of freedom are ($\alpha$, $\beta$, $\gamma$). b) Schematic of lab setup. Two counter-propagating 1550\,nm laser beams form the optical trap inside a vacuum chamber. The intensity of the light is controlled by an acousto-optic modulator (AOM) and the polarization state by an electro-optic modulator (EOM) and polarization control paddles (PCP). The motion of levitated nanoparticles is detected by balanced detectors for the $z$-, $\alpha$- and $\beta$-motion and a quadrant photodetector (QPD) for the $x, y$-motion. These detection signals are sent to a data acquisition unit, lock-in amplifiers, and a computer with an FPGA module for data recording. A 532\,nm pulsed laser is used for LIAD loading of the nanorods into the optical trap \cite{nikkhou2021}. c) A sample PSD of a $700\times200\,\unit{\nm}$ nanorod trapped at a total optical power of $300\,\unit{\mW}$, at pressure $22\,\unit{\mbar}$, which is fitted by \cref{eq:PSD} shown as the black line.}
\label{fig:theory}
\end{figure*}

We levitate the nanorods shown in \cref{sec:fabrication} in a $1550\,\unit{\nm}$ counter-propagating optical tweezer, and treat it as a Rayleigh particle. Considering a rigid body, the levitated nanorod has six degrees of freedom, 3 translations ($x, y, z$) and 3 rotations ($\alpha, \beta, \gamma$), as shown \sfref{fig:theory}{a}. In vacuum, these modes are deeply underdamped, with residual gas collisions acting as the dominant source of stochastic thermal noise. In our setup, the rotation about the long symmetry axis $\gamma$ is unconfined by the optical potential due to the cylindrical symmetry, while the remaining modes are confined and measurable. A minimal model for the nanorod dynamics is provided below.

\paragraph*{Trapped dynamics --}
Assuming that the confined modes oscillate / librate about their equilibrium position, the optical potential can be approximated as harmonic for all degrees of freedom and their dynamics described by the Langevin equation
\begin{subequations}\label{eq:langevin}
\begin{align}
   &\ddot q(t) + \Gamma_{q}  \dot  q(t) + \omega_{q}^2  q(t) = \frac{1}{C_q} W_q(t), \\
   &\text{with}\quad C_q =\begin{cases}
   M, & q \in \{x,y,z\} \\
   I, & q \in \{\alpha,\beta\}
   \end{cases}.
\end{align}
\end{subequations}
Here $\Gamma_q$ is the momentum damping rate, $\omega_q=2\pi f_q$ is the oscillation frequency, $M$ is the mass of the nanorod and $I = \frac{1}{12}M(L^2+\frac{3}{4}D^2)$ is the moment of inertia for a nanorod of length $L$ and diameter $D$.
The stochastic thermal noise is modelled as Gaussian white noise with correlation $\langle  W_q(t)  W_q(t')\rangle = 2C_q\Gamma_q k_\mathrm{B}T_{\mathrm{gas}}\delta(t-t')$ with $T_\mathrm{gas}$ the temperature of the gas colliding with the nanoparticle. The corresponding displacement power spectral density (PSD)
    \begin{equation}
        S_{qq}(\omega) = \frac{k_\mathrm{B} T_q}{\pi C_q }\frac{\Gamma_q }{(\omega^2 -\omega_q^2)^2 + \Gamma_q^2\omega^2},
        \label{eq:PSD}
    \end{equation}
which is used fit to experimental data to extract the frequency $\omega_q$ and the damping $\Gamma_q$.

Considering the profile of the trapping beam as Gaussian, the frequencies for all modes are given by \cite{kuhn2017full}
\begin{subequations}\label{eq:freq}
\begin{align}
    f_x &= \frac{1}{2\pi} \sqrt{\frac{8P_{\mathrm{opt}}\chi_{\parallel}}{\pi\rho c w_x^3 w_y}},  \quad  f_y = \frac{1}{2\pi} \sqrt{\frac{8P_{\mathrm{opt}}\chi_{\parallel}}{\pi\rho c w_x w_y^3}}, \\
     f_z &= \frac{1}{2\pi} \sqrt{\frac{4k^2P_{\mathrm{opt}}\chi_{\parallel}}{\pi\rho c w_x w_y}}, \quad f_{\alpha} = \frac{1}{2\pi} \sqrt{\frac{48P_{\mathrm{opt}}\Delta\chi}{\pi\rho c w_x w_y L^2}}, \\
    f_{\beta} &= \frac{1}{2\pi} \sqrt{\frac{48P_{\mathrm{opt}}\chi_{\parallel}}{\pi\rho c w_x w_y L^2} \left(\frac{\Delta\chi}{\chi_{\parallel}} + \frac{(kL)^2}{12}\right)},
\end{align}
\end{subequations}
where $\rho = 2330\,\unit{kg/m^3}$  is the density of the rod, $P_{\mathrm{opt}}$ and $k$ are the power and wavenumber of the trapping beam, and $w_0$ is the beam waist in the transverse axes $x$ and $y$.
The anisotropic geometry of the nanorod results in directional susceptibility to the polarization of the light field $\chi_{\parallel}$, $\chi_{\perp}$, with susceptibility anisotropy $\Delta \chi = \chi_{\parallel} - \chi_{\perp}$. This leads to the alignment of the nanorod along the optical polarization axis (the $y-$axis). The susceptibility is approximated by considering an infinitely long cylinder \cite{kuhn2017full}: 
\begin{equation}\label{equ:susceptibility}
    \chi_{\parallel} = \epsilon_\mathrm{r} - 1,\qquad 
    \chi_{\perp} = \frac{2(\epsilon_{\mathrm{r}} - 1)}{\epsilon_{\mathrm{r}} + 1}.
\end{equation}
 The enhanced susceptibility of an aligned cylinder as compared to a spherical object greatly increases the optical trap depth for a given optical power, thereby offering a route to reduced phase and photon shot noise \cite{Jain_2016,Meyer_2019}.

At pressures for which the gas mean free path is much larger than the nanoparticle size, the residual-gas damping rate is linearly proportional to pressure. In this regime, the damping rates on a cylinder are given by \cite{Martinetz_2018,millen2019single}
\begin{subequations}\label{equ:momentum_damping}
    \begin{align}
         &\frac{\Gamma_{x,z}}{2 \pi} = A(P_{\mathrm{gas}}) \eta_T\left(2 + c_{\mathrm{acc}} \left(-\frac{1}{2} + \frac{\pi}{4} + \frac{R}{L}\right)\right), \\
         &\frac{\Gamma_{y}}{2 \pi} = A(P_{\mathrm{gas}}) \eta_T \left(4\frac{R}{L} + c_{\mathrm{acc}} \left(1 - 2\frac{R}{L} + \frac{\pi}{2}\frac{R}{L}\right)\right), \\
         \intertext{for the translational degrees, and}
         &\frac{\Gamma_{\alpha, \beta}}{2 \pi} = A(P_{\mathrm{gas}}) \eta_T \frac{L^2}{3R^2 + L^2} \Bigg[2 + 12\frac{R^3}{L^3} \label{equ:alpha_damping}\\ 
         & + c_{\mathrm{acc}} \left(-\frac{1}{2} + \frac{\pi}{4} + 3\frac{R}{L} + 6\frac{R^2}{L^2} + \left(\frac{3\pi}{2}-6\right)\frac{R^3}{L^3}\right)\Bigg],  \notag
    \end{align}
\end{subequations}
for the rotational degrees. Here $D=2R$ , the prefactor $A(P_{\mathrm{gas}})= \frac{3\sqrt{2}}{2\pi}\frac{\mu_v\sigma_{\mathrm{gas}}}{k_{\mathrm B}T_{\mathrm{gas}}\rho}\frac{P_{\mathrm{gas}}}{D}$ with $P_{\mathrm{gas}}$ the gas pressure and $\mu_v =\frac{2\sqrt{m_\mathrm{gas}k_BT_\mathrm{gas}}}{3\sqrt{\pi}\sigma_\mathrm{gas}}$ the viscosity coefficient. The collision cross-section is given by $\sigma_{\mathrm{gas}} = \pi d_{\mathrm{m}}^2$ with the diameter of air molecules is $d_{\mathrm{m}} = 0.372\,\unit{nm}$. The phenomenological hot-emission correction $\eta_T=1+\frac{1}{16}\sqrt{\frac{T_\mathrm{em}}{T_{\mathrm{gas}}}}$ with $T_{\mathrm{gas}}$ and $T_{\mathrm{em}}=c_{\mathrm{acc}}(T_{\mathrm{int}}-T_{\mathrm{gas}})+T_{\mathrm{gas}}$ the temperatures of the gas molecules before and after colliding with the nanorod, and $T_{\mathrm{int}}$ the internal temperature of the rod.
The accommodation coefficient $c_{\mathrm{acc}}$ describes the fraction of thermal energy removed from the surface of the nanorod via collisions with the gas molecules, with $c_{\mathrm{acc}} = 1$ indicating the gas molecules thermalising with the nanorod. This coefficient is extracted from experimental data fitted by \cref{eq:PSD}.

\paragraph*{Rotational dynamics --}
Rotation of the nanorod in the $\alpha$ degree of freedom can be driven in our experiment by changing the trapping light polarization from linear to circular in the transverse plane. For circular polarization, the optical potential for $\alpha$ becomes approximately flat, while the radiation pressure exerts a constant torque on the nanorod, which reads \cite{Stickler_2016,kuhn2017full}
\begin{subequations}
    \begin{align}
        &N_\alpha = \frac{P_\mathrm{opt} \Delta \chi L^2 D^4 k^3}{48cw_0^2} [\Delta\chi \eta_1(kL) + \chi_\perp \eta_2 (kL)], \\
        \intertext{with the correction terms}
        &\eta_1(kl) = \frac{3}{4} \int_{-1}^1 (1- \xi^2)\,\mathrm{sinc}^2\left(\frac{kL\xi}{2}\right) d\xi, \\
        &\eta_2(kl) = \frac{3}{8} \int_{-1}^1 (1- 3\xi^2)\,\mathrm{sinc}^2\left(\frac{kL\xi}{2}\right) d\xi.
    \end{align}
\end{subequations}
For short nanorods, $kL\ll1$, the generalized Rayleigh--Gans approximation gives $\eta_1\simeq1$ and $\eta_2\simeq0$. The maximum steady-state rotation frequency of the nanorod is obtained by balancing the optical torque with the gas damping rate \cite{kuhn2017full}
\begin{equation}\label{eq:rotf}
    f^\mathrm{rot}_{\alpha, \mathrm{max}} = \frac{N_\alpha}{2\pi I \Gamma_\alpha}.
\end{equation}

\paragraph*{Experimental setup --}

This experiment is designed to study the ro-translational dynamics of levitated nanorods, which requires precise control of the polarization states of both beams that form the optical trap. This control is achieved using a high speed fibre electro-optic polarization controller (EOspace) and  polarization control paddles (PCPs), as shown in \sfref{fig:theory}{b}. To drive rotation in the $\alpha$-direction in a counter-propagating trap requires opposite polarization handedness in the two beams, achieved by introducing a $\pi/2$ phase shift in one beam via mirror reflection. Rotation in the $\beta$-direction is not studied in this work, but can be achieved by generating transverse orbital angular momentum \cite{Hu_2023}. 

Laser-induced acoustic desorption (LIAD) \cite{nikkhou2021} is used to load a nanorod into the optical trap, at pressures of a few mbar. The particle is captured in a linearly polarized trap using lenses with numerical aperture ${\rm NA}=0.43$. Since we trap the nanoparticles in a standing wave, there are multiple nodes at which they could be captured. We move the particles to the highest intensity part of the optical field by translating one of the optical fiber out-couplers, but the process is imperfect. In Section \cref{sec:syst} we describe how we extract the optical beam waist from the data.

Balanced photodetectors and a quadrant photodiode (QPD) are used to detect the five degrees of freedom of the nanorod \cite{kuhn2017full} (see \sfref{fig:theory}{b}). The $x$- and $y$-motions are measured by sending a portion of the light transmitted through the trap to the QPD. The $z$-motion is measured by subtracting a reference beam from light transmitted through the trap with a balanced photodetector. The $\alpha$-motion is measured via the birefringence of the nanorod by performing a balanced polarization resolved measurement on the light transmitted through the trap. The $\beta$-motion is similarly measured via a polarization resolved measurement on light scattered sideways from the nanorod. We note that in our counter-propagating beam system, light transmitted through the trap contains light both forward- and back-scattered from the nanorod.  

Switching the trap to circular polarization drives the rotation of the nanorod in the $\alpha$ direction. For low-frequency rotation ($\leq 1\,\unit{\MHz}$), the $\alpha$-motion is detected using the same polarization-sensitive balanced detection as described above. For high-frequency $\alpha$-rotation, a $200\,\unit{\MHz}$ Photoreceiver (FEMTO OE-300-IN-03 $200 \,\unit{\MHz}$ Variable Gain Photoreceiver) after the circulator is used instead to resolve fast intensity modulations.

\begin{figure*}[t]
    \centering
    \begin{overpic}[width=\linewidth]{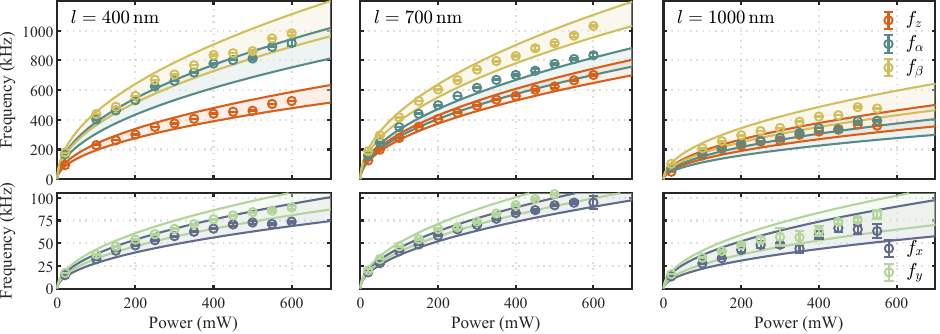}
    \put(0,34){a)}
    \put(36,34){b)}
    \put(68.3,34){c)}
    \end{overpic}
    \caption{Measured trapping frequencies of translational and librational modes (open circles) against total optical power for nanorods of $D = 200$\,nm and $L =$ (a) $400\,\unit{\nm}$, (b) $700\,\unit{\nm}$ and (c) $1000\,\unit{\nm}$. For each nanorod, the beam waist is calculated by fitting to \cref{equ:beam_waist_from_ratio}, and then then the susceptibility can be calculated using \cref{eq:freq}. The shaded region represents the model in \cref{eq:freq} taking into account uncertainty in the parameters.}
    \label{fig:bwfittotal}
\end{figure*}

\section{Full characterization of levitated nanorods with different geometries}\label{sec:syst}

We now explore the veracity of these models for describing the behaviour of levitated silicon nanorods with various lengths and diameters. 

\paragraph*{Trapping frequencies --} 
The motion power spectral densities (PSDs) of a $700\times200\,\unit{\nm}$ nanorod trapped at a pressure of $22\,\unit{\mbar}$ with total optical power $300\,\unit{\mW}$ is shown in \sfref{fig:theory}{c}. The separation in $x$, $y$ frequencies arises from the anisotropic transverse geometry of a focussed Gaussian beam. The black curves represent the fitting of each peak using \cref{eq:PSD}, from which the frequency $f_q$ and momentum damping $\Gamma_q$ for $q\in\{x,y,z,\alpha,\beta\}$ are extracted.

In \cref{fig:bwfittotal} we record the variation in trapping frequency with total optical power, for nanorods of dimension $400 \times 200\,\unit{\nm}$, $700 \times 200\,\unit{\nm}$, and $1000 \times 200\,\unit{\nm}$. These data follow the scaling relation $f_q\propto\sqrt{P}$ as predicted by \cref{eq:freq}.  The transverse beam waists can be extracted from the frequency ratios according to
\begin{equation}\label{equ:beam_waist_from_ratio}
    w_{i} = \frac{\sqrt{2}}{k}\frac{f_z}{f_{i}},
\end{equation}
with $i = x, y$. The extracted values are reported in \cref{tab:susceptibilities}. The beam waists vary due to the nanorods being trapped at different axial positions in the optical standing wave. Averaged beam waists for all measurements are $w_{x}^{\rm ave} = (2.24\pm 0.28)\,\unit{\um}$ and $w_{y}^{\rm ave} = (1.95\pm 0.27)\,\unit{\um}$. Using the extracted beam waists, the susceptibilities of different nanorods can be subsequently determined, with $\chi_\parallel$ obtained from $f_z$ and $\Delta\chi$  from $f_\beta^2-f_z^2$, as reported in \cref{tab:susceptibilities}. 

These values are significantly different from the predictions in \cref{equ:susceptibility} ($\chi_{\parallel} = 11.1. \Delta\chi = 9.4$, suggesting that the infinite cylinder is a poor model for our nanorods, though these values are in better agreement than considering a prolate spheroid \cite{venermo2005}. We further note that extracting $\Delta\chi$ from $f_\alpha$ yields unphysical values, likely due to an enhanced libration angle, see \cref{apd:sec:libration_angle} for a discussion. The extracted beam-waist and susceptibility values, along with their uncertainties, are used in \cref{eq:freq} and compared to experimental data in \cref{fig:bwfittotal}. 

\begin{table}[t]
    \centering
    \begin{tabular}{|c|cccc|}
        \hline
        Nanorod & $w_x$ ($\unit{\um}$) & $w_y$ ($\unit{\um}$) & $\chi_{\parallel}$ & $\Delta\chi$ \\
        \hline
        $400$--$200$  & $2.31 \pm 0.11$ & $2.00 \pm 0.06$ & $2.61 \pm 0.10$ & $1.50 \pm 0.12$ \\
        $700$--$200$  & $2.44 \pm 0.07$ & $2.19 \pm 0.08$ & $4.08 \pm 0.08$ & $3.71 \pm 0.15$ \\
        $1000$--$200$ & $1.97 \pm 0.18$ & $1.68 \pm 0.10$ & $1.79 \pm 0.14$ & $1.54 \pm 0.56$ \\
        \hline
    \end{tabular}
    \caption{Extracted beam waists and susceptibilities for different nanorods geometries. We note that the susceptibilities do not all agree with the model in \cref{equ:susceptibility}.}
    \label{tab:susceptibilities}
\end{table}

\begin{figure*}[t]
\begin{overpic}[width=\linewidth]{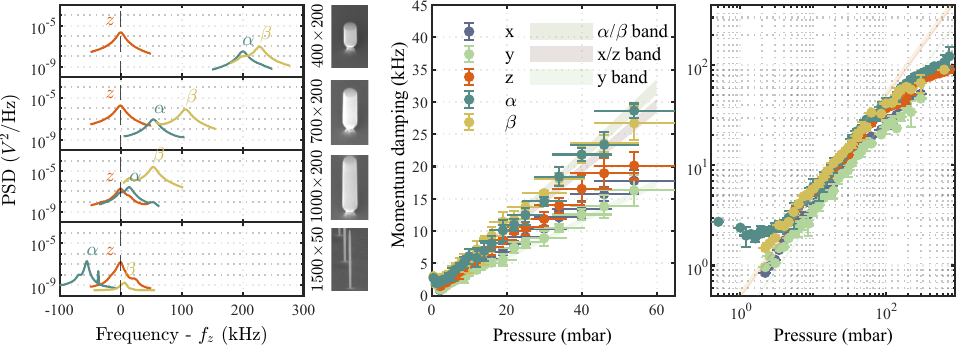}
    \put(0,34.5){a)}
    \put(40.5,34.5){b)}
\end{overpic}
    \caption{a) PSDs of the translational mode $z$ and angular modes $\alpha$ and $\beta$ for nanorods of different dimensions. The frequency axis is referenced to the $z$-mode frequency. b) Momentum damping rates of a $700\times200\,\unit{\nm}$ nanorod in the low-pressure linear regime (left panel) and over the full measured pressure range (right panel). The shaded bands represent the theoretical predictions, with the band widths accounting for uncertainty in the nanorod surface temperature $T_{\rm int}$, which we allow to range from room temperature $298\,\unit{\K}$ to the melting temperature of silicon $1687\,\unit{\K}$.}
    \label{fig:PSDdemo}
\end{figure*}

We demonstrate the ability to tune the relative frequencies of the different trapping modes via fabrication in \sfref{fig:PSDdemo}{a}. As clear from \cref{eq:freq}, the $\beta$-mode always has a larger frequency than that of the $\alpha$-mode. The difference between librational modes and the $z-$modes are the greatest for shortest rod, since the trapping frequencies of the librational degrees of freedom $\alpha$, $\beta$ are inversely proportional to the nanorod length $L$. The $z$-mode frequency surpasses the $\alpha$-mode frequency for nanorods longer than $\sim1000\,\unit{\nm}$, and it can surpass the $\beta$-mode frequency for nanorods longer than $\sim1500\,\unit{\nm}$.

\paragraph*{Momentum damping rate --} The momentum damping rate of each mode can be extracted by fitting the corresponding PSD with \cref{eq:PSD}. As an example, we take a nanorod of size $700\times200\,\unit{\nm}$ and present the momentum damping rates of all modes over a pressure range of 60\,mbar in \sfref{fig:PSDdemo}{b)}. 

We compare the experimental data with the theoretical prediction given by \cref{equ:momentum_damping}, where the damping rate is predicted to be linearly proportional to pressure below $\sim50\,\unit{\mbar}$ (the free molecular flow regime), in the left panel of \sfref{fig:PSDdemo}{b)}. The shaded region of the theoretical prediction reflects the uncertainty in the internal temperature of the nanorod, which we allow to range across the maximum possible range; from room temperature $300\,\unit{\K}$ to the melting temperature of silicon $1687\,\unit{\K}$.The measured damping rates of the $y$, $\alpha$ and $\beta$ modes agree well with the theoretical predictions, whereas the agreement is weaker for the $x$ and $z$ modes. This discrepancy could indicate mixing between the measured damping rates of the translational modes. While the theory predicts the momentum damping along and perpendicular to the nanorod's long axis in the body frame, the experimental measurement is performed along the $x$ and $z$ directions of the lab frame [cf. \cref{fig:theory}]. Thus, the measurement and the theory would only agree if the nanorod were perfectly aligned with the polarization $y$-direction at all time. We investigate the libration angle in the $\alpha$-direction in \cref{apd:sec:libration_angle}, where the analysis suggests a value $>20\unit{\degree}$. We thus conjecture that large libration angles contribute to the mixing of the measured momentum damping rates in the $x$ and $z$ modes.

The measured momentum damping rate for the same nanorod geometry over a wider pressure range is shown in the right panel of \sfref{fig:PSDdemo}{b)} on a doubly logarithmic scale. We observe that the momentum damping plateaus in the high-pressure limit where the mean free path of gas molecules becomes closer to, or smaller than, the nanorod size such that the model in \cref{equ:momentum_damping} no longer holds. We present a physically motivated empirical model in \cref{apd:sec:momentum_damping}.  The momentum damping rate also appears to level-out in the low-pressure limit, which we attribute to instabilities in the nanoparticle motion..

\begin{figure*}[t]
    \centering
    \begin{overpic}[width=\linewidth]{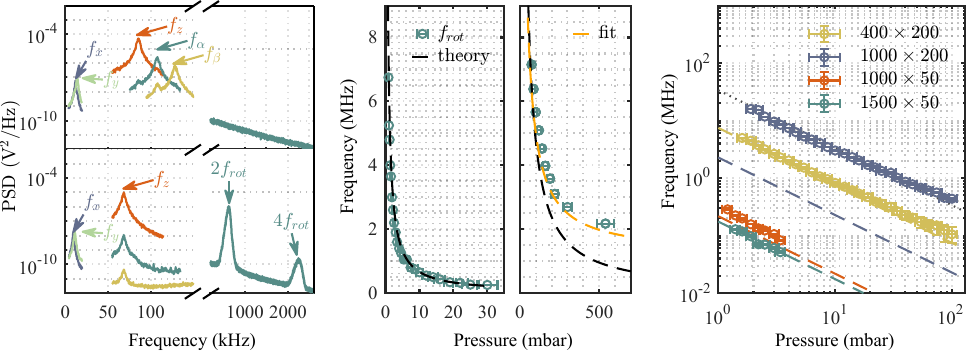}
        \put(0,34){a)}
        \put(34.5,34){b)}
        \put(67,34){c)}
    \end{overpic}
    \caption{Optically driven rotation of silicon nanorods. a) PSDs of a $700\times200\,\unit{\nm}$ nanorod trapped with linearly (top) and circularly (bottom) polarised light. b) Maximum $\alpha$ rotation frequencies of a $700\times200\,\unit{\nm}$ nanorod, at low pressures with a trapping power of $10\,\unit{\mW}$ (left), and at high pressures with a trapping power of $720\,\unit{\mW}$ (right). The dashed black curve shows the theoretical prediction obtained using \cref{equ:alpha_damping}, while the orange curve shows the prediction obtained using the measured momentum damping rate. c) Maximum rotation frequencies (open circles) of nanorods with different sizes at low pressures with a trapping power of $50\,\unit{\mW}$, together with the corresponding theoretical predictions (dashed lines). The grey dotted line is a direct fit to the $1000\times200\,\unit{\nm}$ nanorod data, not a comparison to the model.}
    \label{fig:spincompare}
\end{figure*}

\paragraph*{Optically driven rotation --} 
By switching the optical polarization from linear to circular, the nanorod undergoes rotation in the $\alpha$ direction. A comparison between the frequency spectrum of a nanorod trapped with linear and circular polarization is shown in \sfref{fig:spincompare}{a}. The nanorod dimension is $700\times200\,\unit{\nm}$, the total optical power is $10\,\unit{\mW}$, and the pressure is $7.4\,\unit{\mbar}$. When the polarization is linear, five modes are recorded (upper panel). For circular polarization (lower panel), broad $f_\alpha$ peaks correspond to the (non-resonant) driven rotation. The detectors record $2f_{\rm rot}$ and $4f_{\rm rot}$ due to the inversion symmetry of a cylinder. The $f_{\alpha}$ peak is absent as this is the direction of rotation, and the $f_{\beta}$ peak is gyroscopically stabilized hence also not visible. The $x$, $y$, and $z$ resonance frequencies decrease by $2.75\,\unit{kHz}$, $3\,\unit{kHz}$, and $18.5\,\unit{kHz}$, respectively. This frequency change is explained by the susceptibility changing from $\chi_\parallel$ for a trapped nanorod to an average over both $\chi_\parallel$ and $\chi_\perp$ for a rotating one \cite{kuhn2017full}. As a result, the libration frequencies of translational modes are reduced by around $23\,\%$ when the cylinder is rotating.

For a spinning nanorod, the maximum rotational frequency is pressure dependent through the momentum-damping rate \cite{kuhn2017full}, as described by \cref{eq:rotf}. Taking a nanorod of $700\times200\,\unit{\nm}$ as an example, we compare the measured rotational frequency with the theoretical prediction as a function of pressure in \sfref{fig:spincompare}{b)}. 
The left panel shows good agreement between measurement and theory at pressures below $30\,\unit{mbar}$, while the right panel shows a disagreement at pressures above $\sim50\,\unit{mbar}$. The disagreement at high pressures comes from the fact that the momentum-damping rate can no longer be described by \cref{equ:alpha_damping} [cf. \sfref{fig:PSDdemo}{b}]. To account for this, we fit the high-pressure momentum-damping rate using an empirical model described in \cref{apd:sec:momentum_damping}. We use this model to correct \cref{eq:rotf}, and show the results as the orange dashed curve in \sfref{fig:spincompare}{b)}. This illustrates that we are able to describe the rotational frequencies even at near-ambient pressures using our models.


We explore the rotation rates of a range of nanorod geometries in \sfref{fig:spincompare}{c)}. The theory agrees well with most of the measurements, with the exception of $1000\times200\,\unit{\nm}$ nanorods. We attribute this discrepancy to the fact that for unknown reasons this geometry of nanorod cannot be trapped at the trap centre, which results in lower-than-expected motional frequencies. From the rotation data, we can estimate $\Delta\chi_{1000-200}^{\rm rot}\approx6$, shown by the gray dotted line.
We note that data for $D = 50\,\unit{\nm}$ nanorods are collected only up to $3\,\unit{\mbar}$ since the torque is inadequate to drive rotation above this pressure.

\begin{figure}[t]
    \centering
    \begin{overpic}[width=\linewidth]{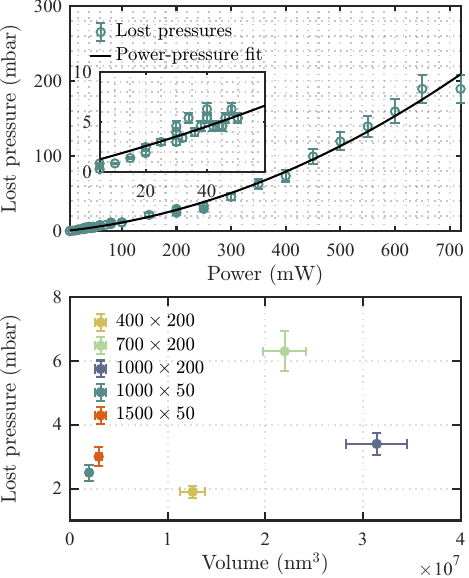}
        \put(0,98){a)}
        \put(0,48){b)}
    \end{overpic}
    \caption{Investigation of nanorod trapping instability. a) Lost pressure for $700\times200\,\unit{\nm}$ nanorods under tweezer powers from $5\,\unit{\mW}$ to $720\,\unit{\mW}$. The data are fitted with a second-order polynomial. b) Lost pressure against the volume of different nanorods.}
    \label{fig:lostp-V}
\end{figure}

\section{Trapping instability}\label{sec:inst}
The instability of optically levitated nanoparticles at low pressures is a known issue \cite{Millen2014,Hoang2016_loss,Pi2025,alavi_2025}. Consistent with previous observations, we find that nanorods escape from the optical trap when the pressure is reduced below a critical pressure.
Here we report this loss pressure against the trapping power for $700\times200\,\unit{\nm}$ nanorods in \sfref{fig:lostp-V}{a}. We are able to levitate a nanorod at a trapping power as low as $5\,\unit{\mW}$ and reduce the pressure to $0.3\,\unit{mbar}$ before the particle is lost. Given no obvious change in the nanorod's PSD as the pressure is reduced, we conclude that the instability is unlikely to be caused by structural damage of the rod. The fitted data show that the loss pressure increases with trapping power, and is approximately linearly dependent on power in the low-power regime. As suggested by previous works \cite{Millen2014,Hoang2016_loss}, this behaviour suggests that the instability results from a balance between absorption-induced heating and cooling through collisions with the background gas.

Loss pressures against the size of the nanorods are reported in \sfref{fig:lostp-V}{b}. Here all nanorods are levitated with total optical trapping power of $50\,\unit{\mW}$. In general, we see that larger particle has a higher loss pressure, due to an increased trap depth, but the relation is not strong.

We compare our work to the work in ref \cite{Junnemann_2025}, which reports runaway heating of silicon nanospheres. We note that for the same optical intensity our silicon nanorods are stable to a significantly lower pressure, especially surprising since the volume of the cylinders is approximately twice that used in the cited study. Due to the shape-enhanced susceptibility we are able to trap below 1mbar when using low powers. We note that the emissive properties of a cylinder may be significantly different to that of a sphere, and that the cited study used colloidally produces silicon which may have different absorptive properties. We believe that there are complex dynamic effects which also effect the stability of the silicon nanoparticles, worthy of study beyond this preprint.

\section{Conclusions}

In this work, we demonstrated the optical levitation of nanofabricated silicon nanorods with aspect ratios ranging from 2 to 30. The fabrication process produces highly uniform particles with diameters as small as $50\,\unit{\nm}$ and lengths up to $1500\,\unit{\nm}$, providing a reproducible platform for studying the translational and rotational dynamics of anisotropic nanoparticles. 

We characterized the frequency--power relations for nanorods with different aspect ratios and demonstrated librational-mode frequencies above $1\,\unit{\MHz}$ at high trapping powers. The measured frequency--power relations follow the predicted $f\propto\sqrt{P}$ scaling, enabling us to extract the effective beam waists and optical susceptibilities of nanorods with different dimensions.
We also characterized the pressure-dependent momentum-damping rates of all nanorod modes. In the low-pressure regime, the measured momentum-damping rates of several modes agree with the free-molecular gas-damping model. Deviations observed for some translational modes are likely associated with the large libration angles of the nanorods and the resulting mixing of motion between the particle and laboratory frames. At higher pressures, the damping departs from the free-molecular approximation and approaches a pressure-independent regime, consistent with the general theory of gas damping.

By switching the trapping light from linear to circular polarization, a large optical torque can be applied and drive the nanorods into rapid rotation. The measured maximum rotation frequencies are well described by the balance between optical torque and gas damping when the appropriate pressure-dependent damping rate is used.

Finally, we investigated the instability that limits levitation at low pressures. We find that the loss pressure increases with trapping power and is approximately linearly dependent on power in the low-power regime. This instability could be governed by the competition between absorption-induced heating and cooling through collisions with the residual gas and by blackbody emission. Further investigation of the nanorod internal temperature is required for a more comprehensive addressing of this instability problem.

The combination of high geometric uniformity, megahertz librational frequencies, and strong optical torque makes levitated silicon nanorods promising systems for precision torque sensing and for studies of rotational dynamics far from equilibrium. With improved control of absorption and low-pressure stability, these particles could provide a route towards quantum control of angular motion and the generation and detection of angular-momentum superposition states.

\begin{acknowledgments}
This project has received funding from the European Research Council (ERC) under the European Union Horizon 2020 research and innovation programme (grant agreement nos. 803277 \& 957463). We also acknowledge the Engineering and Physical Sciences Research Council International Quantum Technologies Network Grant EP/W02683X/1. We appreciate many discussions with Professor Benjamin Stickler.
\end{acknowledgments}

\bibliographystyle{apsrev4-2}
\bibliography{refs.bib}

\newpage
\appendix
\renewcommand{\thefigure}{\Alph{section}\arabic{figure}}
\setcounter{figure}{0}

\makeatletter
\@addtoreset{figure}{section}
\makeatother

\section{Details for the theoretical model}
Suppose the nanorod is cylindrical and its angular motions are described by the Euler angles $(\alpha,\beta,\gamma)$, with the principle moments of inertia $I_\alpha = I_\beta = I$ and $\gamma$-axis is along the rod [cf. \sfref{fig:theory}{a}]. The Hamiltonian of the levitated nanorod is $H = T + V$ with the kinetic energy reads
\begin{equation}
    T = \frac{\mathbf{p}_{\rm cm}^2}{2m} + \frac{1}{2I}\left[\frac{(p_\alpha - p_\gamma\cos\beta)^2}{\sin^2\beta} + p_\beta^2\right] + \frac{p_\gamma^2}{2I_\gamma},
\end{equation}
where $\bf p_{\rm cm}$ is the centre-of-mass momentum and $m$ is the mass of the particle.
Suppose the tweezer beam has a Gaussian profile, the potential reads
\begin{align}
    V & = - \frac{\epsilon_0 V_{rod} \chi_{\parallel}E_0^2}{4} \left(\frac{w_0}{w(z)}\right)^2 \exp\left(\frac{-2 x^2}{w_0^2}\right)\exp\left(\frac{-2 y^2}{w_0^2}\right) \notag\\
    & \times\left\{\frac{1}{2} + \frac{1}{2}\cos(2kz)S\left[\vec{m},\vec{e}_z\right] \right\}\left[ \frac{\chi_\perp}{\chi_\parallel} + \frac{\Delta \chi}{\chi_\parallel}(\vec{m}\cdot\vec{e}_x)^2\right],
\end{align}
where we assume the particle's orientation is $\vec{m}$, the beam is propagating along $z$ direction and the polarization is along the $x$ direction. $S[\vec{m},\vec{e}_z] = \sin(k l \vec{m}\cdot\vec{e}_z)/ k l \vec{m}\cdot\vec{e}_z$ is the shape function accounting for the particle's finite extension \cite{Stickler_2016}. The Euler angles give $\vec{m}\cdot\vec{e}_z=\cos(\beta)$ and $\vec{m}\cdot\vec{e}_x = \sin{\beta}\cos\alpha$.
$w(z)$ is the the beam waist radius at position $z$ and $w_0 = w(0)$. The particle is trapped at the beam centre where $w(z) \approx w_0$. Notice that if the particle moves along the beam to position $z$, effectively the trapping capability weakens due to the beam being less focused. 
Assuming small libration angles, the potential is approximated to be 
\begin{align}
        V & = - \frac{\epsilon_0 V_{rod} \chi_{\parallel}E_0^2}{4} \exp\left(\frac{-2 x^2}{w_x^2}\right)\exp\left(\frac{-2 y^2}{w_y^2}\right) \cos(kz)^2 \notag \\
        &\hspace{1cm}\times\left[ \frac{\chi_\perp}{\chi_\parallel} + \frac{\Delta \chi}{\chi_\parallel}\sin^2\beta\cos^2\alpha\right].
\end{align}
Here $\bm{\chi} = (\chi_\perp,\chi_\perp,\chi_\parallel)$ are the susceptibilities perpendicular or along the nanorod.
Given the relation between the laser power and the electric field strength $E_0^2 = \frac{8 P_t}{\pi c \varepsilon_0 w_x w_y}$, the potential gives the corresponding frequencies of all modes in \cref{eq:freq}, where the extra term in $\omega_\beta$ is from the accounting of particle's finite extension.

\section{Libration angle}\label{apd:sec:libration_angle}

\begin{figure}[t]
\begin{overpic}[width=\linewidth]{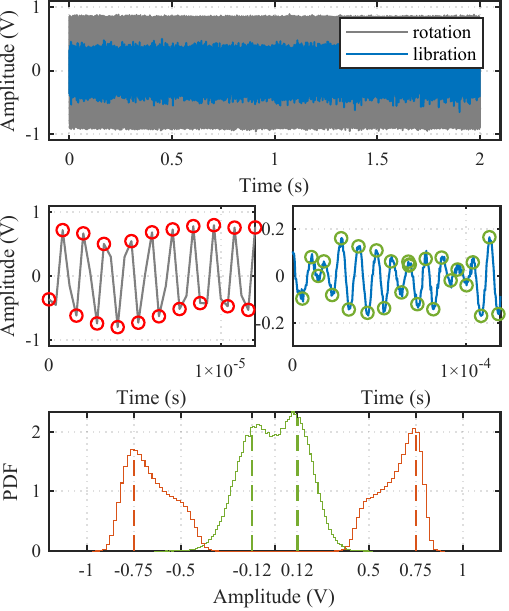}
    \put(0,99){a)}
    \put(0,66){b)}
    \put(0,32){c)}
\end{overpic}
    \caption{Libration and rotation data for the $\alpha$-mode. a) Detector signals for both motions. b) The turning points of signals for the first $10\,\unit{\us}$ rotation data (red circles) and the first $100\,\unit{\us}$ libration data (green circles). c) Histograms for the collected turning points. The data record is $2\,\unit{\s}$ long, with the nanorod librating in the $\alpha$-mode at $103\,\unit{\kHz}$ and rotating at $790\,\unit{\kHz}$.}
    \label{apd:fig:Libration_angle}
\end{figure}

We use the same balanced detector to measure the $\alpha$ mode as the beam polarization changes from linear to circular. Representative time traces of $\alpha$-mode libration and rotation are shown in \sfref{apd:fig:Libration_angle}{a}. Assuming the detector response remains unchanged between the two polarization configurations, the measured signal amplitude corresponds to the transmitted-light intensity induced by the alignment of the nanorod relative to the beam polarization, with zero representing the nanorod being aligned with the beam polarization.
We model the signal as $S(\alpha) = \zeta I_{\rm max} \sin(\alpha)^2$, with $\zeta$ the converting factor, $I_{\rm max}$ is the maximal light intensity when the nanorod is perpendicular to the beam polarization, and $\alpha$ the libration (rotation) angle.
The turning points of the signal correspond to extrema of the libration or rotation angles, which are collected by post-processing the data as shown in \sfref{apd:fig:Libration_angle}{b} and their histograms in \sfref{apd:fig:Libration_angle}{c}. The ratio between the libration signal and the maximum rotation signal provides an estimate of the libration angle, $\alpha = \arcsin{\sqrt{S^{\rm lib}/S^{\rm rot}_{\rm max}}}$. From the rotation data, we obtain the maximal rotation signal of approximately $S^{\rm rot}_{\rm max} = 0.75\,\unit{\V}$. The libration signal statistics are determined from the extracted extrema, \textit{e.g.} $S^{\rm lib}_{\rm+ mean} = 0.12\,\unit{\V}$. we estimate that the central $50\%$ of the libration-angle distribution lies within $|\alpha|\leq 23\unit{\degree}$, while the central $90\%$ lies within $|\alpha|\leq 34\unit{\degree}$. There is $10\%$ of the measured extrema correspond to libration angles exceeding $34\unit{\degree}$. These results indicate that the $\alpha$-mode dynamics extend beyond the linear-angle approximation.

\section{The fitting of momentum damping}\label{apd:sec:momentum_damping}
\begin{figure}
    \centering
    \includegraphics[width=\linewidth]{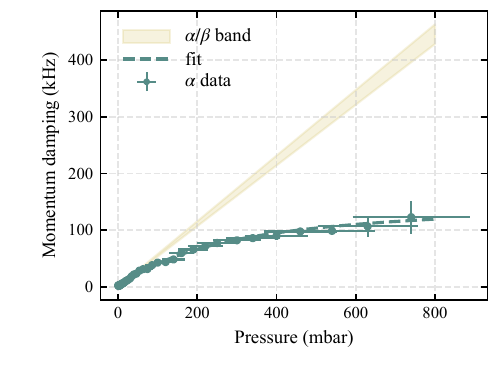}
    \caption{Fitting the momentum damping rate of $\alpha$-mode at high pressures with \cref{apd:equ:momentum_damping_Kn}.}
    \label{apd:fig:gamma_fit}
\end{figure}
The momentum damping rate due to gas collision to a levitated nanosphere is given by \cite{Beresnev_Chernyak_Fomyagin_1990,millen2019single}
\begin{equation}\label{apd:equ:momentum_damping_Kn}
    \frac{\Gamma_{\rm gas}}{2\pi} {=}  \frac{3 \mu_v a}{m} \frac{0.619}{0.619+{\rm Kn}}\left(1+\frac{0.31\,{\rm Kn}}{0.785+1.152\,{\rm Kn}+{\rm Kn}^2}\right),
\end{equation}
where $\mu_v=\mu_v =\frac{2\sqrt{m_\mathrm{gas}k_BT_\mathrm{gas}}}{3\sqrt{\pi}\sigma_\mathrm{gas}}$ is the viscosity of the residual gas with $\sigma_{\rm gas} = \pi d_{\rm m}^2$, $d_{\rm m}=0.372\,\unit{\nm}$ the diameter and $m_{\rm gas}$ the mass of the air molecule, and ${\rm Kn} = \bar{l}/a$ is the Knudsen number for the free mean path $\bar{l} = k_BT_\mathrm{gas}/(\sqrt{2}\sigma_{\rm gas}P_{\rm gas})$. $a$ and $m$ are the radius and the mass of the nanosphere. At low pressures, the mean free path is much larger than the radius of the particle (${\rm Kn}\gg 1$), and the momentum damping rate depends linearly on the gas pressure $P_{\rm gas}$, as stated in \cref{equ:momentum_damping}. At high pressures (${\rm Kn}\ll 1$), the momentum damping rate becomes independent of pressure, in accordance with Stokes’ law.

To extract the momentum damping of the $\alpha$ mode at high pressures, we fit the data with
\cref{apd:equ:momentum_damping_Kn} as shown in \cref{apd:fig:gamma_fit}. Note that the fit assumes a spherical particle of radius $a_{\rm fit} = 327\,\unit{\nm}$.

\end{document}